\documentclass[prl,twocolumn,showpacs,preprintnumbers,amsmath,amssymb,superscriptaddress]{revtex4}

\usepackage{graphicx}
\usepackage{dcolumn}
\usepackage{bm}
\usepackage{color} 

\begin{document}

\preprint{Manuscript}

\title{Rounding of Phase Transitions in Cylindrical Pores}

\author{Dorothea Wilms}
\author{Alexander Winkler}
\author{Peter Virnau}
\author{Kurt Binder}
\affiliation{Institute of Physics, Johannes Gutenberg University Mainz,\\
Staudingerweg 7, D-55128 Mainz, Germany}



\begin{abstract}
Phase transitions of systems confined in long cylindrical pores
(capillary condensation, wetting, crystallization, etc.)
are intrinsically not sharply defined but rounded. The
finite size of the cross section causes destruction of long range order along the
pore axis by spontaneous nucleation of domain walls.
This rounding is analyzed for two models (Ising/lattice gas and
Asakura-Oosawa model for colloid-polymer mixtures)
by Monte Carlo simulations and interpreted by a phenomenological
theory. We show that characteristic differences between the
behavior of pores of finite length and infinitely long pores
occur. In pores of finite length a rounded transition
occurs first, from phase coexistence between two states towards a
multi-domain configuration. A second transition to the axially homogeneous phase follows near pore criticality.
\end{abstract}

\maketitle Fluids and fluid mixtures in nano- and microporous
materials (pore diameters from 1 nm to 150 nm) play
important roles in various industries (extracting oil and gas from
porous rocks; use as catalysts or for mixture separation in the
chemical and pharmaceutical industry; nanofluidic devices, etc.)
\cite{1,2,3}. The interplay of finite size and surface effects
strongly modifies the phase behavior of such confined fluids 
\cite{1,3,4,5,6,7,8,9,10,11,12,13,14,15,16,18,19,20} in
comparison with the bulk. The vapor to liquid transition is
shifted (``capillary condensation''), as well as critical points
\cite{3,4,9,12}. Effects of wetting \cite{21} on phase coexistence give rise to interesting patterns (plugs versus
capsules versus tube structures etc. \cite{7}). However, although
various phase diagrams (different from the bulk) have been
proposed (e.g.~\cite{1,3,7,9,12,13,18}), many aspects hitherto are
not well understood. E.g., the ``critical point'' where
adsorption/desorption hysteresis vanishes seems to be
systematically lower than the critical temperature where the
density difference between the vapor-like and liquid-like states
vanishes \cite{12}, in contrast to what theories have predicted
\cite{14}.

However, a crucial aspect (stressed only in a few pioneering
studies \cite{3,8}, and in the context of Ising/lattice gas models
\cite{22,23,24}) is the rounding of all transitions,
caused by the quasi-one-dimensional character of a fluid in a long
cylindrical pore with cross-sectional radius $R$. With the
current progress of producing pores of well-controlled diameter
varying from the nanoscale (carbon nanotubes \cite{24,25,26})
to arrays of pores in silicon wafers \cite{27}, up to 150
nm wide and of well-controlled length, experiments become feasible
which are not plagued by effects of random disorder, which occur
in porous glasses \cite{1,28}. Thus, it is important to
understand the phase transitions in pores more
precisely, considering both the radius $R$ and the length $L$ of
the pore as variables (the important role of $L$ has so far been
largely disregarded). In the present Letter, we
elucidate the rounding of vapor-liquid type transitions in cylindrical pores, based on Monte Carlo simulations of two
generic models and a phenomenological theory. We show
that, even in the absence of precursors of wetting, two
rounded transitions occur. Near the pore critical temperature at the pressure where vapor and liquid in the pore coexist, a
rounded transition occurs from a axially homogenous state to a
multi-domain configuration, where vapor-like and liquid-like
domains alternate. The properties in this region depend
strongly on $R$ but not at all on $L$. In contrast, at lower
temperatures the system makes a transition, where the
full capillary is either in a vapor-like or a liquid-like state.
The location of this transition depends on $L$, and the vapor to
liquid transition is accompanied by a pronounced hysteresis. We
also show that the effective (size-dependent) free energy exhibits
well-defined spinodals (as a finite size effect), but they do not
control dynamics. Nucleation of domain walls becomes dominant
when their free energy cost are small (of order of a few $k_BT$,
 $T$ being the temperature; henceforth $k_B=1$). This domain wall nucleation explains why the hysteresis disappears
far below the capillary critical region for small pores.

\begin{figure}[tbp]
\includegraphics[scale=0.5]{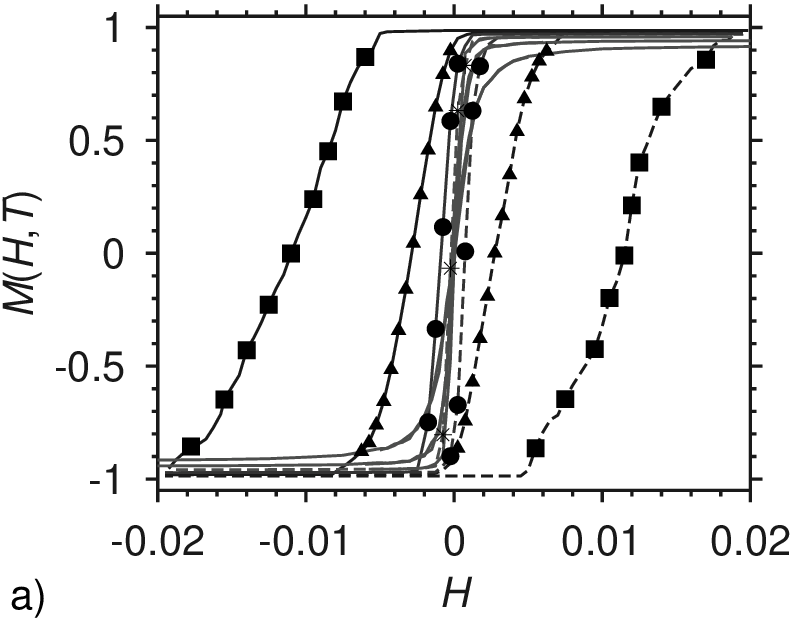}
\includegraphics[scale=0.5]{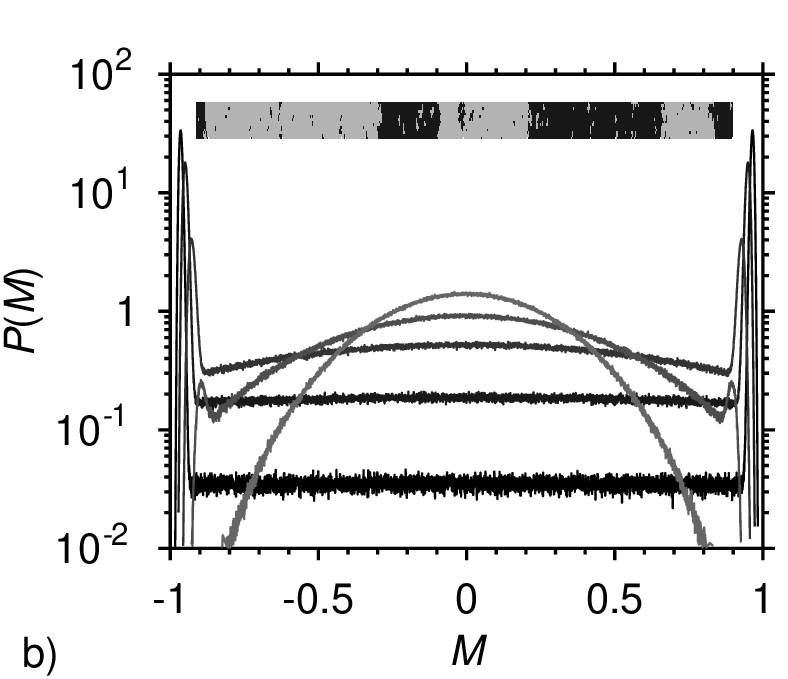}
\caption{(a) Magnetization of Ising strips for $L=480$,
$D=10$ plotted vs. field $H$ at $T=1.5 \blacksquare, 1.6 \blacktriangle, 1.7 \bullet, 1.8 \star, 1.9$ and $2.0$.
Runs with decreasing $H$ are shown as full curves, with increasing
$H$ as broken curves. A detailed analysis shows that the hysteresis
disappears at $T_0=1.9$ in this case. (b) Distribution $P(M)$ vs.
$M$ for $H=0$ and $T=1.8 - 2.2$ from bottom to top at $M=0$. The inset is a typical 
snapshot at $T=2.1$ containing multiple domains stretched in y-direction by a factor $\approx 4$.}\label{fig1}
\end{figure}

The simplest model that already shows some of these effects is the
two-dimensional (2-d) Ising model on the square lattice in the geometry
of $D \times L$ strips with periodic boundary conditions in both
directions \cite{22,23,24,30}. While this 2-d model lacks the surface effects due to the walls of real 3-d pores, it exhibits already the disappearance of hysteresis far below pore criticality, since the condition $L>>D$ suffices to stabilize multi-domain states, as we will show below. Spins $S_i= \pm 1$ at lattice sites
$i$ interact with their nearest neighbors with an energy
$J=1$, and an external field $H$. We apply the standard single
spin-flip Monte Carlo algorithm \cite{31} and record the
magnetization $M = \sum\limits_{i=1}^N S_i/N$ ($N=LD$; the lattice
spacing being the unit of length) as a function of $H$ at various
$T$. We start out with all spins up and $H=0.05$. The
system runs for a ``time" $t= 2\cdot10^6$ Monte Carlo steps per spin (MCS).
Then we decrease $H$ in steps of $\Delta H = 0.001$, 
and run the simulation at each field for the same time, until we reach
$H=-0.05$. Afterwards, we reverse the process and increase the field
stepwise by $\Delta H$ until we are back at $H=0.05$. The width
of the resulting hysteresis loops (Fig.~\ref{fig1}a) strongly
decreases with increasing $T$ and for $T > T_0(L,D)$, the ``hysteresis critical point'', a hysteresis is no longer
observed. However, when we record the probability distribution $P(M)$ for $H=0$ with the Wolff
cluster algorithm \cite{31,32} we observe that $P(M)$ still
exhibits peaks very close to the (exactly known \cite{33})
spontaneous magnetization $M_s$ at temperatures $T > T_0 (L,D)$.
While for $L=480$, $D=10$ these peaks can be followed up to about
$T=2.1$, the peaks occur up to about the critical
temperature for $D=10$ and smaller $L$, e.g. $L=40$. However, 
at $T_0(L,D)$ an important change also occurs in $P(M)$: while for $T<T_0 (L,D)$ for a
wide range of $M$ $P(M)$ is strictly independent of $M$
(corresponding to a slab configuration which contains exactly two
non-interacting interfaces \cite{34}), for $T>T_0(L,D)$ a third
broad peak in $P(M)$ appears at $M=0$. An examination of snapshot
pictures of the system (Fig.~\ref{fig1}b) reveals that this 3$^{rd}$ peak is due to
multi-domain configurations \cite{30,35,36}.

\begin{figure}[tbp]
\includegraphics[scale=0.5]{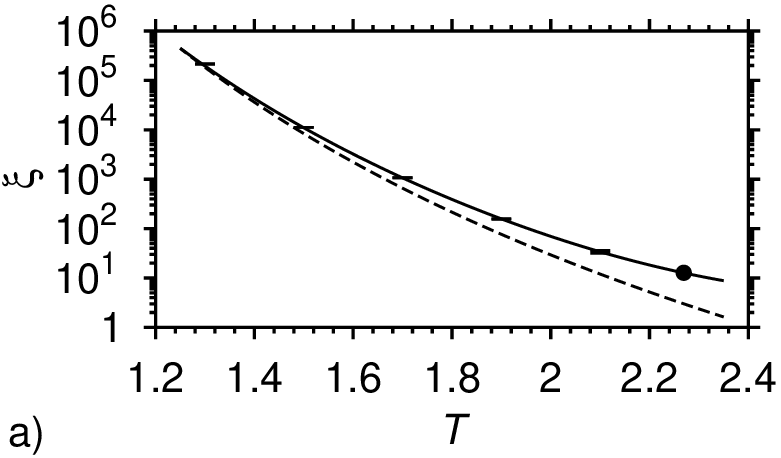}
\includegraphics[scale=0.5]{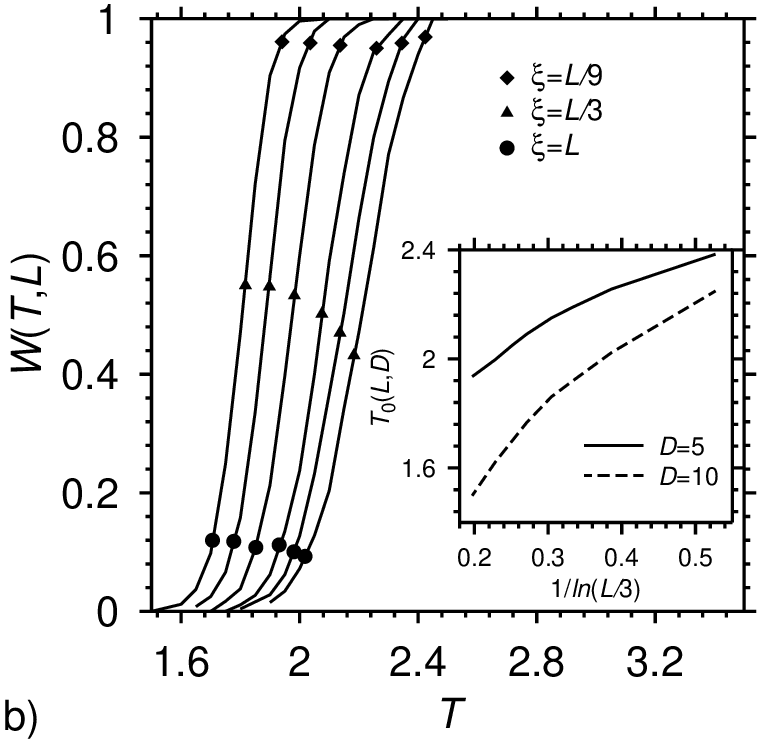}
\caption{(a) Correlation length $\xi$ (on a logarithmic scale) plotted
vs. $T$ for Ising strips of width $D=10$. Monte Carlo results
(shown with error bars) were extracted for a system of $L=10000$,
recording the wavevector-dependent susceptibility $\chi(\vec{k})=N
\langle |M(\vec{k})|^2\rangle$ where $M(\vec{k})=\sum S_j \exp (i
\vec{k} \cdot \vec{r}_j)$ for $\vec{k}$ oriented in the long
direction and $k=k_{\rm min}=2 \pi/L$, using the formula quoted in
the text.
 Transfer matrix results were computed from the exact formula \{Eq.(4.39)of
 \cite{23}\} for the $D \times \infty$ system. Broken curve shows the
 approximation $\xi \approx \exp (D \sigma/T)$ where $\sigma$ is the
 exactly known interfacial tension of the Ising model. The value
 of $\xi$ at $T_c$ \cite{22}, $\xi_c=(4D/\pi)$, is shown as a dot. (b) 
Weight $W$ of the central peak for strips of width $D=10$
 plotted vs. temperature for $L=1000, 480, 240, 120, 80, 60$ from the left to the right. The
 symbols indicate $\xi=L$, $L/3$ and $L/9$,
 respectively. Insert shows a plot of $T_0(L,D)$ vs. $L$ (note logarithmic scale)
 for two choices of $D$.}\label{fig2}
\end{figure}
Such multi-domain configurations can in fact be predicted when one
computes the correlation length $\xi$ along the strip
(Fig.~\ref{fig2}a) by transfer matrix methods (for $L \rightarrow
\infty$ \cite{23}) or Monte Carlo (for $L \gg \xi$
\cite{35,36}). The latter estimates were extracted from the wave
vector-dependent susceptibility $\chi(\vec{k})$ for $\vec{k} =
\vec{k}_{\rm min} =(2 \pi/L,0)$
\begin{equation} \label{eq1}
\xi= \frac{1}{2 \sin (k_{\rm min}/2)} \Big[\frac{\chi
(0)}{\chi(\vec{k}_{\rm min})} -1 \Big]^{1/2} \quad ,
\end{equation}

and agree perfectly with the exact results. Thus, for very long strips the statistical 
errors are also well under
control. 

This correlation length below criticality (where well-defined 
domains exist) just measures the typical distance between domain walls along the strip. 
The approximation based on the (exactly known \cite{37}) interfacial 
free energy $\sigma$, $\xi \approx\exp (D \sigma/T)$ becomes only accurate when $\xi \geq10^5$, i.e. at temperatures
much lower than those of interest for Fig.~\ref{fig1}.
This
 simply represents the well-known argument
\cite{38} that long-range order in quasi-one-dimensional systems
is destroyed due to the entropy gain of putting interfaces into the system. The
free energy difference (relative to the single-domain state) for a
state with $n$ (non-interacting) interfaces is $F=nF_{\rm int} + n
T \ln (n/eL)$, where the total free energy cost of one interface
is given by $F_{\rm int}=D \sigma$.

The occurrence of the central peak at $T$ near $T_0 (L,D)$ means
that when $T$ is raised at $H=0$ there is a transition from
nonzero $\langle |M|\rangle$ for $T<T_0(L,D)$ to a
state with no order $(\langle |M|\rangle \ll M_s)$ for $T >
T_0(L,D)$. We characterize this transition by the
weight of the central peak of $P(M)$, defined as
$W=\int\limits_{-m}^{+m} P(M) dM/\int\limits^{+1}_{-1} P(M) dM$
where $\pm m $ are the locations of the minima of $P(M)$.
Fig.~\ref{fig2} shows that the ``equal weight'' rule (first order
transitions from one state to another state occur when the weights
of the two states are $W=1/2$) roughly corresponds to the condition
$\xi \approx L/3$. With
increasing $L$ the transition gets shifted to lower temperature
and also gets sharper. Since $W \approx 0.1$ for $\xi=L$ and $W
\approx 0.9$ for $\xi=L/9$, we use $\xi \approx\exp (D \sigma/T) \approx\exp (2D/T)$ for low $T$ for $L
\rightarrow \infty$ to estimate both the location of the
transition and its width $\Delta T$,
\begin{eqnarray} \label{eq3}
&& T_0(L,D)\approx \frac{2D}{\ln (L/3)},\ \ 
L \rightarrow \infty,\ \ 
\frac{\Delta T}{T_0(L,D)} \approx \frac{\ln 3}{\ln (L/9)}
\end{eqnarray}

\begin{figure}[tbp] 
\includegraphics[scale=0.5]{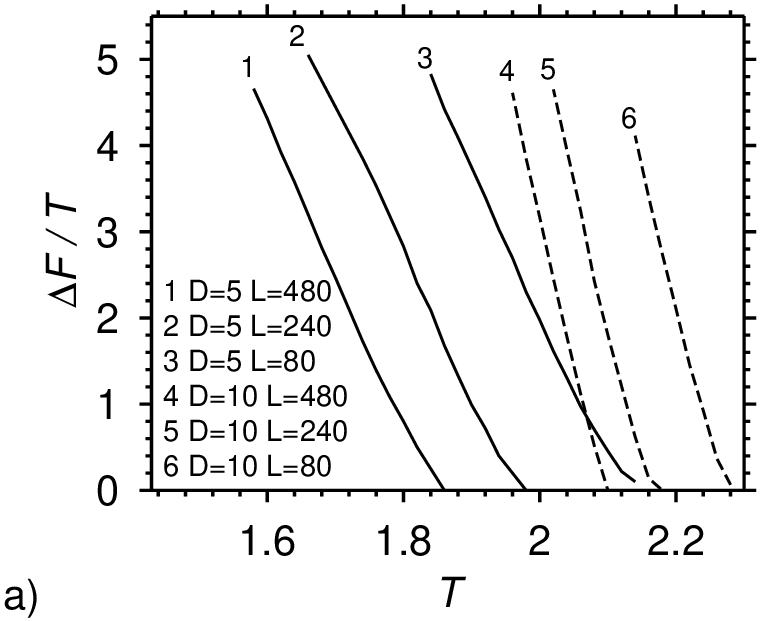}
\includegraphics[scale=0.5]{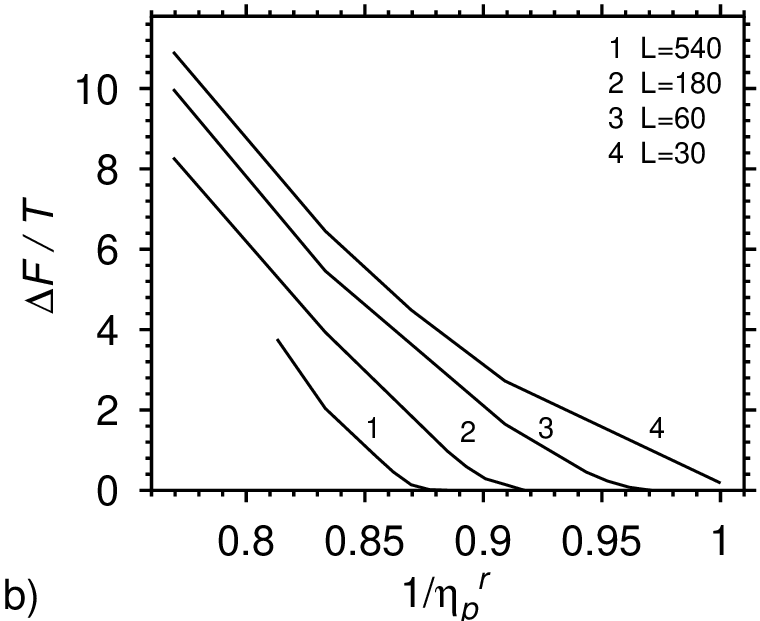}
\caption[font=smallest]{(a) Barrier $\Delta F / T$ against nucleation of interfaces
in Ising strips plotted vs. $T$. Several choices of $L$ and $D$
are shown, as indicated. (b) Barrier $\Delta F / T$ against nucleation
of interfaces in the AO model confined to cylindrical pores of
diameter $D=12$ plotted vs. inverse polymer reservoir packing
fraction $1/\eta_p^r$. }\label{fig3}
\end{figure}
Finally, defining
a barrier $\Delta F$ from $P(M)$ as $\Delta F=T \ln [P(M_{\rm
max})/P(m)]$ we see (Fig.~\ref{fig3}a) why the transition at
$T_0(L,D)$ is related to the vanishing of hysteresis: $\Delta F(T)
\rightarrow 0$ at a temperature where $M_{\rm max}$ and $m$ merge,
which is close to $T_0 + \Delta T$ where $W \rightarrow 1$.
Actually, the hysteresis already vanishes when $\Delta F/T \approx
10$ since then nucleation of interfaces is sufficiently easy.

In order to show that these results
carry over to fluids confined in long cylindrical pores, we have
studied the Asakura-Oosawa (AO) model \cite{39} of
colloid-polymer-mixtures. The latter system is attractive for
experiments: the large colloid size renders effects of the
atomistic corrugation of pore walls negligible, and facilitates
observation of wetting layers and interfaces \cite{40}. We
describe colloids as hard spheres of radius $r_c=1$, and
polymers as soft spheres of radius $r_p=0.8$.
Polymer-colloid overlap (as well as colloid-colloid overlap) is
strictly forbidden, while polymers can overlap with no energy
cost. The phase diagram of this model in the bulk and for thin
films has already been carefully
studied \cite{41,42}. At the cylinder radius $R=D/2$ we apply a hard
wall, which may overlap with neither colloids nor polymers. This
boundary condition at the surface leads to
an entropic attraction of the colloidal particles to the wall
\cite{42}, causing the formation of a precursor of a wetting layer
(a true wetting layer can only form in the
limit $D \rightarrow \infty$, of course \cite{42}).

\begin{figure}[tbp] 
\includegraphics[scale=0.5]{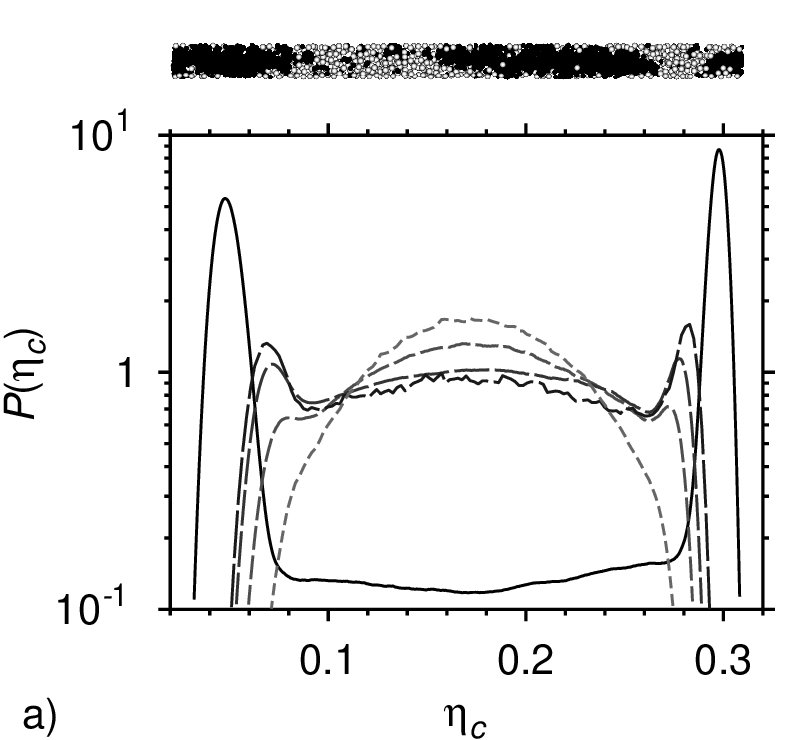}
\includegraphics[scale=0.5]{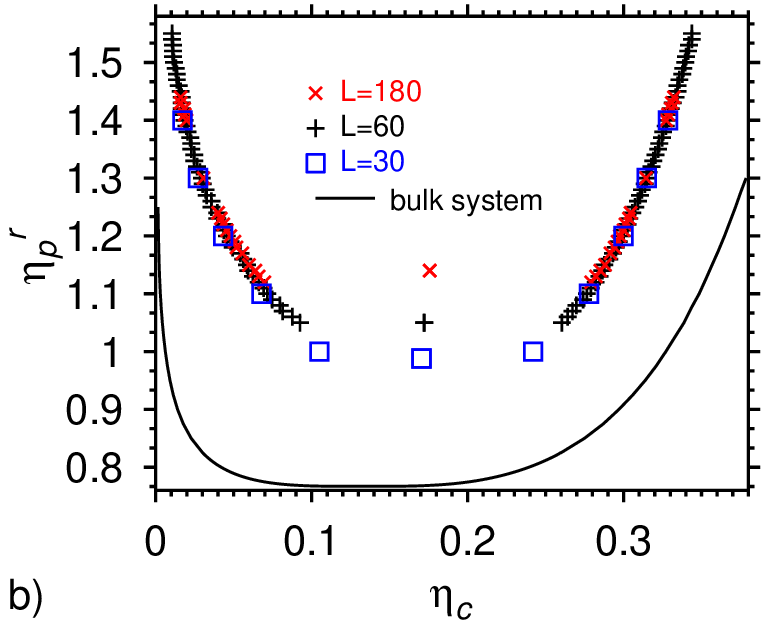}
\caption[smallest]{(a) Distribution $P(\eta_c)$ of the number of colloids in a
cylinder of diameter $D=12$ and length $L=180$ (all lengths are
measured in units of the colloid radius) for $\eta^r_p=1.075, 1.10, 1.115, 1.118, 1.20$
from top to bottom at $\langle\eta_c\rangle \approx 0.175$. 
Above the plot we show a typical snapshot (cross section through the cylinder) 
at $\eta^r_p=1.10$ containing multiple domains. 
(b) Phase diagram of the AO model in a cylindrical
pore of diameter $D=12$ and lengths $L=30$, 60 and 180, as indicated.
The full curve is the bulk coexistence curve \cite{41}. Note that
the points shown near $\langle\eta_c\rangle \approx 0.16$ to $0.17$ mark
$\eta^r_{p0}(D,L)$ for three choices of $L$. }\label{fig4}
\end{figure}
For this model, the polymer fugacity $\exp (\mu_p/k_BT)$ or the
related ``polymer reservoir packing fraction'' $\eta_p^r =(4 \pi
r_p^3/3) \exp (\mu_p/k_BT)$ plays the role of an inverse
temperature like variable, while the colloid packing fraction
$\eta_c=(4 \pi r_c^3/3)N_c/V$ ($N_c$ is the number of
colloids in the system of volume $V=\pi R^2L)$ is the
order parameter density. Fig.~\ref{fig4}(a), as a
counterpart of Fig.~\ref{fig1}(b), shows $P(\eta_c)$ for various
values of $\eta^r_p$. (The same Grand Canonical 
Monte Carlo methods as in \cite{41} 
are used.) One can clearly distinguish the crossover from an
(asymmetric) double-peak distribution to a structure with three
peaks, and finally a single peak, which only narrows when
$\eta^r_p$ is close to $\eta^r_{p,crit}=0.766$ \cite{41}. The
``phase diagram'', where the coexisting polymer-rich and
colloid-rich phases are estimated from the
left-most to the right-most peak in Fig.~\ref{fig4}(a), is shown in
Fig.~\ref{fig4}(b). Fig.~\ref{fig3}(b) shows that the 
barrier against nucleation of interfaces across the pore strongly decreases 
with increasing $L$, and we have checked \cite{36} that hysteresis disappears 
when the barrier is a few $k_B T$, as for the Ising model.

In summary, we have clarified the nature of phase coexistence between 
vapor and liquid phases of fluids (or fluid-fluid phase coexistence of 
mixtures) in long cylindrical pores, depending on pore length $L$ and pore 
diameter $D$. While at high temperatures the structure of the fluid is axially 
symmetric, phase separation in axial direction sets in at the 
coexistence pressure when the correlation length (of the density 
fluctuations) $\xi$ grows to the order of $D$. Below the pore 
critical temperature $\xi$ measures the distance between domain walls, and at 
a much lower temperature (where $\xi \approx L/3$) a second (again rounded) 
transition occurs (the pore then is either in an axially homogeneous vapor-like 
or liquid-like state). The onset of adsorption hysteresis in the capillary 
is linked to this lower transition. A wetting transition (possible at a flat 
surface of a semi-infinite system) is also expected to be strongly rounded in 
narrow pores, and should not change the above conclusions. Our findings 
provide insight to understand experiments and simulations of fluids in 
pores, explaining the existence of a ``hysteresis critical point'' distinct 
from the pore critical point. A prediction 
that experiments could test is the decrease of the hysteresis critical 
temperature with increasing pore length.

Acknowledgments: We are grateful to the Deutsche Forschungsgemeinschaft (DFG) for 
support (grants No. TR6/A5,C4) and to the NIC Juelich for a generous grant of computing time.


\begin{thebibliography}{99}
\bibitem{1} L.D. Gelb et al.,
Rep. Progr. Phys. {\bf 62}, 1573 (1999)

\bibitem{2} T. Thorsen et al., Science
 {\bf 298}, 580 (2002)

 \bibitem{3} I. Brovchenko and A. Oleinikova, {\it Interface and Confined
 Water} (Elsevier, Amsterdam, 2008)

 \bibitem{4} R. Evans et al., J. Chem.
 Soc. Faraday Trans {\bf 2}, 1763 (1986)

\bibitem{5} G. Heffelfinger et al., Mol. Phys.
{\bf 60}, 1381 (1987)

\bibitem{6} R. Evans, J. Phys.: Condens. Matter {\bf 46}, 9899 (1990)

\bibitem{7} A.J. Liu et al.,
Phys. Rev. Lett. {\bf 65}, 1897 (1990)

\bibitem{8} L.D. Gelb and K.E. Gubbins, Phys. Rev. E{\bf 56}, 3185 (1997)

\bibitem{9} K. Morishige et al., Langmuir {\bf 13},
3494 (1997)

\bibitem{10} P.L. Ravikovich et al.,
J. Phys. Chem. B{\bf101}, 3671 (1997)

\bibitem{11} A.V. Neimark et al.,
J. Colloid Interface Sci. {\bf 207}, 159 (1998)

\bibitem{12} K. Morishige and M. Shikimi, J. Chem. Phys. {\bf108}, 7821 (1998)

\bibitem{13} W.D. Machin, Langmuir {\bf15}, 169 (1999)

\bibitem{14} A. Vishnyakov and A.V. Neimark, J. Phys. Chem. B{\bf105}, 7009
(2001)

\bibitem{15} H.K. Christenson, J. Phys.: Condens. Matter {\bf13}, R95 (2001)

\bibitem{16} A. Schreiber et al., Mol. Phys.
{\bf 100}, 2097 (2002); K.G. Kornev et al., Adv. Coll. Interface
Sci. {\bf 96}, 143 (2002)

\bibitem{18} J. Hoffmann and P. Nielaba, Phys. Rev. E{\bf67}, 036115 (2003)

\bibitem{19} C. Alba-Simionesco et al. J. Phys.: Condens. Matter {\bf 18}, R15 (2006)

\bibitem{20} I. Brovchenko, A. Geiger, and A. Oleinikova, J. Phys. Condens. Matter
 {\bf 16}, S5345 (2007)

 \bibitem{21} S. Dietrich, in {\it Phase Transitions and Critical Phenomena},
 edited by C. Domb and J.L. Lebowitz (Academic, London, 1988) Vol
 12, Chap. 1

 \bibitem{22} V. Privman and M.E. Fisher, J. Stat. Phys. {\bf 33}, 385 (1983)

 \bibitem{23} M.N. Barber, in {\it Phase Transition and Critical
 Phenomena} edited by C. Domb and J.L. Lebowitz (Academic, London,
 1983) Vol 8, Chap. 2.

 \bibitem{24} P. Nowakowski and M. Napiorkowski, J. Phys. A: Math. Theor.
 {\bf 42}, 475005 (2009)

 \bibitem{25} S. Inoue et al., J. Phys. Chem. B{\bf102}, 4689 (1998)

 \bibitem{26} M. Meyyappan (ed.) {\it Carbon Nanotubes: Science and Applications}
 (CRC Press, Boca Raton, 2004)

 \bibitem{27} Z.N. Yu et al., J. Vac. Sci.
 Technol. B{\bf21}, 2874 (2003)

 \bibitem{28} P. Wiltzius et al., Phys. Rev. Lett. {\bf 62}, 804 (1989); M.Y. Lin et al.,
 Phys. Rev. Lett.  {\bf 72}, 2207 (1994)

\bibitem{30} E.V. Albano et al.,
Z. Phys. B: Condens. Matter {\bf77}, 445 (1989)

\bibitem{31} D.P. Landau and K. Binder, {\it A Guide to Monte Carlo Simulation in Statistical
Physics, 3$^{rd}$ ed.} (Cambridge Univ. Press, Cambridge, 2009)

\bibitem{32} U. Wolff, Phys. Rev. Lett. {\bf62}, 361 (1989)

\bibitem{33} C.N. Yang, Phys. Rev. {\bf85}, 808 (1952)

\bibitem{34} K. Binder, Phys. Rev. A{\bf25}, 1699 (1982)

\bibitem{35} D. Wilms, Diplomarbeit, Johannes Gutenberg Univ. Mainz (2009, unpublished)

\bibitem{36} More details will be presented in a forthcoming full
paper (A. Winkler et al., unpublished)

\bibitem{37} L. Onsager, Phys. Rev. {\bf65}, 117 (1944)

\bibitem{38} L.D. Landau and E.M. Lifshitz, {\it Statistical
Physics, 3$^{rd}$ ed.} (Pergamon Press, Oxford, 1959)

\bibitem{39} S. Asakura and F. Oosawa, J. Chem. Phys. {\bf22},
1255 (1957)

\bibitem{40} Y. Hennequin et al., Phys. Rev. Lett. {\bf100}, 178305
(2008)

\bibitem{41} R.L.C. Vink and J. Horbach, J. Chem. Phys. {\bf121},
3253 (2004)

\bibitem{42} K. Binder et al., Soft Matter {\bf 4}, 1555 (2008)
\end{thebibliography}
\end{document}